\documentclass[11pt,authoryear]{article}
\usepackage{amsmath, amsthm, amssymb, amsfonts, natbib}
\usepackage{gensymb}
\usepackage{color}
\usepackage{amssymb}
\usepackage{graphicx}
\usepackage{epstopdf}
\usepackage{epsfig,color}
\usepackage{gensymb}
\usepackage{multirow}

\newcommand{\tr}{^{\prime}}
\def\bd#1{\mbox{\boldmath $#1$}}
\newcommand{\diag}{{\rm diag}}

\newcommand{\ot}{\mbox{$\:\otimes \:$}}

\author{Antonio Forcina, Dipartimento di Economia,\\
        Paolo Carbone, Dipartimento di Ingegneria, \\
      University of Perugia, Italy
}
\title{Modelling dark current and hot pixels in imaging sensors}
\begin{document}
\maketitle
\begin{abstract}
A Gaussian mixture model with a complex covariance structure was used to analyse experimental data from images recorded by a digital sensor under darkness, to model the effects of temperature and duration of exposure on artificial signals (dark current), on ordinary and possibly defective (hot) pixels. The model accounts for two components of variance within each latent type: random noise in each image and lack of uniformity within the sensor; both components  are allowed to depend on experimental conditions. The results seem to indicate that the way dark current grows with the duration of exposure and temperature cannot be represented by a simple parametric model. The latent class model detects the presence of at least two types of hot pixels, where the less frequent ones have also a more extreme behaviour. Though the lack of uniformity of the sensor is amplified by duration of exposure and temperature, pixels characteristics seem to deviate in the same direction and with the same relative size.
\end{abstract}
\section{Introduction}
Digital sensors used both in ordinary cameras and in scientific imaging suffer from several anomalies known as dark current, hot pixels and thermal noise, see for instance \cite{Hochedez2014} and \cite{Hochedez2014} for an accurate description based on physical models. In short, dark current denotes a signal which is detected even if no light is hitting the sensor and is known to increase with duration of exposure at a rate depending on temperature. During long exposures, a very small minority of pixels may show a very large signal even in perfect darkness; these are usually called {\it hot pixels} because they show up bright in a dark image; hot pixels may be seen as being occasionally defective; see \cite{Dunlap} for a hypothetical model of hot pixel distribution. In addition, the amount of light recorded by a given pixel is affected by quantization errors due to analog-to-digital conversion and by additional random fluctuations known as {\it thermal noise} effects increasesing with temperature and duration of exposure.

All of these limitations may be almost negligible when imaging daylight scenes where the signal-to-noise ratio is usually large, but they become a serious problem in many applications, like astronomical images where the amount of perceivable light from far away galaxies may be close to that of the background sky.
Anomalies of digital sensors may also impact on other emerging technological areas, such as biometrics, where sensor ageing results in spiky shot noise pixels (see \cite{KaubaUhl}, \cite{Fridrich} and \cite{BergmuellerDebiasi}).
The ordinary way of coping with these problems in astronomical imaging is to obtain an estimate of dark current and hot pixels by averaging a set of dark frames (images taken under the same conditions but in complete darkness) and, in addition, to average several images to reduce noise, see for instance \cite[Chapter 6]{berry2000}.
For an overview of the literature on more sophisticated methods for coping with dark current, hot pixels and noise, see, for instance \cite{Burger2011}, \cite{Chen2015}, \cite{Widenhorn2010}.

This paper is an attempt to formulate a reasonably flexible statistical model that describes the effect of temperature and duration of exposure on dark current, hot pixels and thermal noise. The objective in \cite{Hanselaer2014} is similar, however their models are applied to the overall averages (over pixels and images acquired under constant conditions) as depending on the duration of exposure and temperature. Their approach does not take into account that the effect of experimental conditions on ordinary and hot pixels may be very different. Because of this motivation, here a Gaussian mixture model  \citep{FraRaf02} is used and the issue of how many latent components to consider is examined in some depth; in addition, it is assumed that experimental conditions may affect both the mean and the variance components. Gaussian mixture models have been used in the analysis of dark current by \cite{Svihlik2009}, among others; his objective, however, is to design an algorithm that removes the dark current from a given image and his approach does not seem to be related to the one proposed here. A statistical procedure to detect hot pixels has been proposed in \cite{Leung2009}: essentially, for each pixel, they compute a measure of discrepancy between its signal and the average on a suitable set of its neighbours. Hot pixels are detected by setting a threshold on the distribution of such discrepancies.

Special attention was given to the covariance structure; background knowledge suggests the presence of independent error terms in the value recorded by each pixel in each image together with pixel specific random effects, at least in images taken under the same experimental conditions. A reviewer suggestion that observations taken under different experimental conditions might also be correlated was confirmed by a preliminary exploratory analysis indicating that, at least among ordinary pixels (which account for over 98.5\% of the total) correlations across experimental conditions may be substantial and follow a very specific pattern which will be discussed in detail.

The model was applied to a set of experimental data derived from images taken under complete darkness with a monochromatic Atik 314L+, a good quality CCD (Coupled charged device) designed for astronomical imaging. The results of the analysis, among other things, indicate that: (i) it is possible to detect two categories of hot pixels, where the less frequent ones are also the most deviant; (ii) although the expected values within each latent class increase with duration of exposure and temperature, no simple parametric function seemed to fit adequately; (iii) experimental conditions affect both variance components, but in a different manner.

The paper is organized as follows: after describing the data in Section 2 and the model in Section 3, the results of the analysis are presented in Section 4, followed by a short discussion.
\section{The data}
The experiments were performed on a monochromatic Atik 314L+ CCD camera having an array of 1392 $\times$ 1040 square pixels of 6.45 micron with a 16 bit analog to digital converter. In principle, after taking an image, each pixel can record an integer between 0 and $2^{16}-1$, however the internal processor adds an offset of about 263 to prevent negative values.

Because exposures between 5 and 10 minutes are usual in astronomical imaging, duration of exposures were chosen to be 3, 300 and 600 seconds. Considering that the camera can be cooled up to 27\degree C below ambient temperature, images were taken at -10, 0 and 10 degree centigrade. By combining duration of exposure and temperature, there are 9 different experimental conditions, in addition, a set of images taken at -10\degree C and using the minimum allowed duration were also taken: this should correspond to conditions where dark current can be expected to be almost negligible.

For each of the 10 different experimental conditions, always under complete darkness, a sequence of 30 images were recorded by setting temperature, duration and number of exposures on a specific software. The original data matrix contains 300 images taken on each of the 1,447,680  pixels; because this is a rather large data set for being easily handled on an ordinary computer, a random sample consisting of 100,000 pixels and 10 images under each experimental conditions were selected without replacement to avoid induced correlations. The resulting data may be arranged into a matrix of 100,000 observations and 100 variables.

Formally, for each experimental condition $e$ = $\,\dots 10$, determined by temperature and duration of exposure, $r$ images were recorded in complete darkness under constant conditions on a collection of $n$ pixels which may belong to $K$ different latent types. What is observed is the digitized signal recorded by each pixel in each image.
\section{The statistical model}
\subsection{Exploratory analysis}
It may be expected that observations from the same pixel taken under constant conditions are correlated even within the same latent class because the lack of uniformity in the sensor induces pixel specific random effects; no prior knowledge is available on how these error terms are affected by experimental conditions. For simplicity, one can conceive two main hypothetical models about pixel specific deviations: $H_I$: when experimental conditions are changed, the sensor is re-settled, thus observations should be uncorrelated, or $H_M$ pixels retain their specificities which, however, will be compressed or amplified. Under $H_I$ we expect the correlation matrix to be block diagonal while $H_M$ implies that the covariance matrix within latent class $k$ = $1,\dots ,K$ should have the form
\begin{equation}
\bd \Sigma_k = \diag(\bd\sigma_k^2)\ot \bd I_r + (\bd\tau_k \bd\tau_k\tr) \ot \bd J_r,
\label{CovSt}
\end{equation}
where $\bd I_r$ denotes an identity matrix of size $r$, $\bd J_r$ is a matrix of 1s of the same size, the elements of $\bd\sigma$ measure the image specific uncertainty while those of $\bd\tau$ measure the lack of uniformity in the sensor, in words, if we ignore the image specific error component, the covariance matrix is uniform within replications and has a multiplicative row by column structure across experimental conditions.

An informal assessment can be obtained by inspecting the raw covariance matrix after removing potential hot pixels. The covariance matrix within the collection of pixels whose overall average did not exceeded 287 (99.8\% of the total) was computed and inspected. A small portion of this matrix for four different experimental conditions and two images selected at random within each sequence of 10 is displayed in Table \ref{T:1}
\begin{table}[h]
\caption{ \label{T:1} \it Extract of the raw covariance matrix, four experimental conditions and two randomly selected images within each sequence of 10}
\small
\centering
\begin{tabular}{lcrrrrrrrrr} \\ \hline
 & \hspace{1mm} &
\multicolumn{2}{c}{0.01", -10\degree C} & \multicolumn{2}{c}{3", 0\degree C} &
\multicolumn{2}{c}{300", -10\degree C} & \multicolumn{2}{c}{3", 10\degree C} \\ \hline
\multirow{2}{*}{0.01", -10\degree C} &
  & 267.2 & 1.4   & 1.6 &   1.9 &   2.3 &   3.2 &   7.0 &   7.2 \\
  & &   1.4 & 268.2 & 2.1 &   0.9 &   2.5 &   5.7 &   8.3 &   9.1 \\
\multirow{2}{*}{3", 0\degree C} &
  &   1.6 & 2.1 & 302.0 &   4.1 &   4.7 &   4.0 &  17.9 &  16.8 \\
  & &  1.9 & 0.9 &   4.1 & 304.1 &   4.4 &   4.6 &  19.1 &  18.7 \\
\multirow{2}{*}{300", -10\degree C} &
  &   2.3 & 2.5 &   4.7 &   4.4 & 293.4 &  10.1 &  16.3 &  15.6 \\
  & &  3.2 & 5.7 &   4.0 &   4.6 &  10.1 & 295.9 &  14.9 &  15.0 \\
\multirow{2}{*}{3", 10\degree C} &
  &   7.0 & 8.3 &  17.9 &  19.1 &  16.3 &  14.9 & 471.8 &  82.7 \\
  & & 7.2 & 9.1 &  16.8 &  18.7 &  15.6 &  15.0 &  82.7 & 468.6 \\
\hline
\end{tabular}
\end{table}
The fact that correlations in the $2\times 2$ off diagonal blocks are not substantially different from those in the blocks along the main diagonal, seems to rule out $H_I$.

An informal test of $H_M$ may be based on the following procedure: (i) start from the $100\times 100$ raw covariance matrix and compute the average within each $10\times 10$ block, by ignoring the diagonal elements appearing in the blocks along the main diagonal, (ii) consider the elements of the resulting  $10\times 10$ matrix as a single vector: if $H_M$ was true, the log of this vector would follow a linear model with the row effects equal to the corresponding columns effects and no interaction. The residual variance for this model, once brought back to the original scale, equals 1.21; because averaged covariances are all positive with an overall average of about 16.36, the assumed structure seems to fit reasonably well. Additional evidence  that support $H_M$ will be examined in the final section.
\subsection{Model formulation}
Let $y_{iej}$, $i=1,\dots ,n$, $e=1,\dots, 10$, $j=1,\dots , r$, denote the value recorded by the $i$th pixel under experimental condition $e$ in the $j$th image of a sequence, Though the $y_{iej}$ come from an analogue-to-digital converter and are thus integers, they can actually vary between 0 and $2^{16}-1$, thus they may be seen as almost continuous measurement results. Let $G_k$ denote the assumption that the $i$th pixel belongs to the $k$th latent population,
and suppose that
\begin{equation}
y_{iej} \mid G_k = \mu_{ke} + \sigma_{ke}\epsilon_{kiej} + \tau_{ke}\varepsilon_{ki},
\label{CovS}
\end{equation}
where $\epsilon_{kiej}$ and $\varepsilon_{ki}$ are standard normal variables and are independent for $i=1,\dots ,n$, $e=1,\dots, 10$, $j=1,\dots,r$. Here $\sigma_{ke}$ measures the image specific uncertainty for pixels of type $k$ under experimental conditions $e$ and $\tau_{ke}$ is a measure of the lack of uniformity in the sensor within pixels of type $k$ under experimental conditions $e$. This model has a very important implications concerning the lack of uniformity in the sensor: according to (\ref{CovS}), within a given latent type, experimental conditions can only reduce or amplify the lack of uniformity of the sensor, but each pixel retains the sign and relative size of its deviation from uniformity.

Let $\bd y_{i}$ be the vector with elements $y_{iej}$, with $j$ running faster; the model in (\ref{CovS}) implies that
$$
\bd y_{i}\mid G_k \sim N(\bd\mu_{k}\ot\bd 1_r,\: \bd\Sigma_k).
$$
The model above also assumes that the tendency of a given pixel to belong to latent class $k$ is a feature which does not depend on the experimental conditions, in the sense that temperature and duration may affect the distribution of the measurements but not the latent type.

The dependence of the mean and the variance components on experimental conditions may be formulated in a flexible way within the general assumption that the elements of  $\bd\mu_k,\:\bd\sigma_k,\: \bd\tau_k$, $k=1,\dots K$, can be suitable functions of duration and temperature. To model the dependence of the two variance components on covariates, a log link may be used to ensure that estimates of variance components are non negative, \citep{Aitkin87},
$$
\sigma_{ke}=\exp(\bd z_{e}\tr \bd\alpha_k),\quad \tau_{ke} =\exp(\bd z_{e}\tr \bd\gamma_k),
$$
where $\bd\alpha_k$ and $\bd\gamma_k$ are both of size 3, $\bd z_{e}$ = $(1\: t_e\: d_e)\tr$ with $t_e, \: d_e$ denoting temperature and duration in the $e$-th experiment.

The dependence of $\mu_{ke}$ on temperature and duration of exposure was investigated in more detail by comparing a few parametric models models against a non parametric one. The notion that dark current increases linearly with the duration of exposure is generally accepted in the literature, see for instance \cite{Hochedez2014}, p. 2. The analysis by \cite{Hanselaer2014} seems to support this property, however their conclusions are based on the behaviour of the overall averages, including hot pixels. It is also well known that the rate of growth of dark current increases with temperature. For instance \cite{Hanselaer2014} used a forth degree polynomial which, however, seems to be more an attempt at fitting than interpreting the phenomenon.

The assumption of normality will be submitted to some scrutiny in the actual application where it emerges that the assumption seems to hold with satisfactory accuracy for ordinary pixels but may fail when pixels start to saturate, that is receiving an amount of signal (dark current) close to their maximum capacity. This happens to some degree also to ordinary pixel at 10\degree C and long exposures.
\subsection{The likelihood function and the EM algorithm}
Let $\ell_{ki}$ denote the log-likelihood for the $i$th pixel conditional on $G_k$,
$$
\ell_{ki}=-\frac{1}{2}\left[\log\mid \bd\Sigma_{k}\mid + (\bd y_{i}-\bd\mu_k\ot \bd 1_r)\tr \bd\Sigma_{k}^{-1} (\bd y_{i}-\bd\mu_{k}\ot \bd 1_r)-r\log(\pi) \right],
$$
where the expression for $\bd\Sigma_k$ is given in  (\ref{CovSt}).
Though an explicit inverse of $\bd\Sigma_k$ could, in principle, be derived by symbolic computation, the resulting expression would span several pages of code, while numerical computation requires less than 0.01 seconds on an average computer.

Let $\bd\pi$ be the $K\times 1$ vector of prior probabilities and  $\bd\theta$ be the vector whose elements are the logits of $\bd\pi$ with respect to the first entry. Formally we may write
$$
\bd\pi = \exp(\bd G\bd\theta)/ (\bd 1_K\tr\exp(\bd G\bd\theta)),
$$
where $\bd G$ is an identity matrix without the first column. Under the assumption that observations on different pixels  are independent conditionally on $G_k$, the manifest log likelihood may be written as
$$
L(\bd\theta,\bd\beta,\bd\alpha,\bd\gamma) = \sum_i\log\left[\sum_k \pi_k \exp\left(\ell_{ki}\right) \right].
$$
Let $q_{ki}$ = $\exp(\ell_{ki})$, $\bd q_i$ denote the vector with elements $q_{ki}$; with these notations, we may write $L(\bd \theta,\bd\beta,\bd\alpha,\bd\gamma)$ = $\sum_i\log(\bd\pi\tr\bd q_i)$.

The posterior probabilities that pixel $i$ belongs to latent type $k$ are computed in the E-step as
$$
E_{ki} = \frac{\pi_k q_{ki}}{\sum_k\pi_k q_{ki}}.
$$
Then the complete log likelihood to be maximized in the M-step has the form
$$
L_C(\bd\theta,\bd\beta,\bd\alpha,\bd\gamma) = \sum_i\sum_k E_{ki}\log
\left[\pi_k\bd q_{ki})\right].
$$
Because $L_C$ can be factorized as
$$
L_C(\bd\theta,\bd\beta,\bd\alpha,\bd\gamma) = \sum_k\log(\pi_k)\sum_i E_{ki}+\sum_k \left[\sum_i E_{ki}\ell_{ki}\right],
$$
an explicit estimate may be derived for the prior probabilities: $\hat\pi_k$ = $E_{.k}/E_{..}$; for the second component maximization can be applied separately for each $k$ with respect to the corresponding parameters where the expression to be maximized takes the form $L_{k}$ = $\sum_i E_{ki} \ell_{ki}$.
An expression for the score vector will be made available as supplementary material and exploits known expressions for differentiating the determinant and the inverse of the covariance matrix. The modified Fisher-scoring algorithm of \cite{Forcina16} which uses the empirical information matrix was used and seems to work well in this context.

Because the expected information matrix could not be computed with reasonable accuracy due to numerical problems arising when computing derivatives of the manifest likelihood, a non parametric bootstrap was used to estimate standard errors.
\section{Analysis of the data}
\subsection{Model selection}
If we take for granted the notion that there are ordinary and hot pixels, one could set $K=2$ latent classes. However, for instance, \cite{Leung2009} claim that they detected two different types of hot pixels, with one type behaving in a less discrepant way. A formal procedure for choosing $K$ may be based on different criteria; it has been observed that the usual Bayesian information criteria (BIC) may tend to choose a larger $K$ when the number of observations is large relative to the number of parameters. Table \ref{tab:2}, in addition to the value of BIC, gives also that of ICL proposed by \cite{Biernacki2000}  and NEC, see \cite{Celeux1996}.
\begin{table}[h]
\caption{ \label{tab:2} \it Scaled values of three information criteria for choosing $K$}
\begin{center}
\begin{tabular}{lrrr} \hline
Criteria & $K=2$ & $K=3$ & $K=4$ \\ \hline
BIC/$10^7$ & 4.7225 & 4.7195 & 4.7165 \\
ICL/$10^7$ & 4.7225 & 4.7195 & 4.7165 \\
NEC x $10^7$ & 0.0204 & 9.1996 & 9.2552 \\
 \hline
\end{tabular}
\end{center}
\end{table}
The reason for BIC and ICL to be equal up to the first 5 significant digits is that chnges in the overall entropy in the posterior probabilities is almost negligible relative to the log-likelihood. According to the model with $K=2$, the proportion of hot pixels is about 0.23\%. With $K=3$ the really hot pixels are about 0.19\%, in addition a new category of moderately hot pixels which account for about 1.23\% of the total is detected. By setting $K=4$, the latent type of very hot pixels, is split into two groups with the more extreme one being about 0.03\% of the total. Values of $K$ greater than 4 lead to detect even more extreme collections of hot pixels, but this leads to numerical instabilities. A reasonable compromise, between ICL and NEC seems to choose $K=3$.

A satisfactory model for $\bd\mu_{k}$ was difficult to identify, mainly because of the peculiar behaviour of the two categories of hot pixels; let $d_h,\: h=1,\dots,4$ and $t_l,\: l=1,2,3$ denote, respectively, duration and temperature.
The largest parametric model containing interactions between temperature and duration (LEI) allows the intercept parameter on the log scale to depend on duration
$$
\mu_e=\mu_{hl} = \beta_1+\exp(\beta_2) d_h + exp(\beta_{2+h} + \beta_6  t_l);
$$
this model, which requires 6 parameters within each latent class, was compared against a non parametric model (NPM) which does not impose any functional restriction on the 10 elements of $\bd\mu_{k}$. Because LEI is  nested within NPM, model selection may be based on the likelihood ratio;  this is greater than 70,000 0n 12 degrees of freedom, leading to rejection. The BIC leads to the same conclusion with (2.35 against 2.36)$\times 10^7$.
\subsection{Main results}
\begin{figure}[htb]
\centering
\includegraphics[width=\textwidth,height=4.5cm]{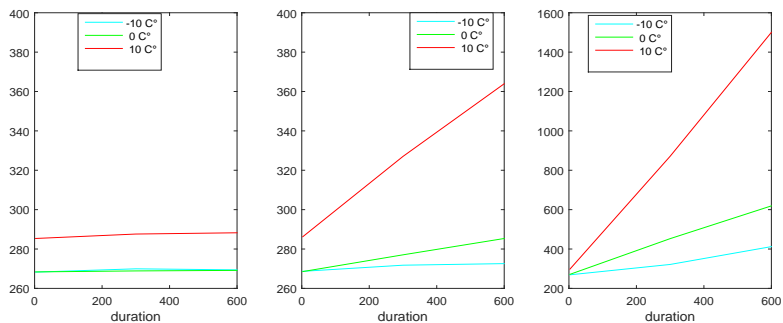}
\caption{\label{Fig1} Mean values (estimated under NPM) versus duration, ordinary pixel in the left panel, moderately hot pixels in the middle panel and very hot pixels in the left panel. }
\end{figure}

The estimated mean values as functions of duration, separately for each temperature, are displayed in Figure\ref{Fig1} with a panel for each latent type. The effect of duration on ordinary pixels is almost negligible and approximately linear among hot pixels. The effect of temperature is more substantial and at 10\degree C the effect can be perceived even among ordinary pixels.

\begin{figure}[htb]
\centering
\includegraphics[width=\textwidth,height=4.5cm]{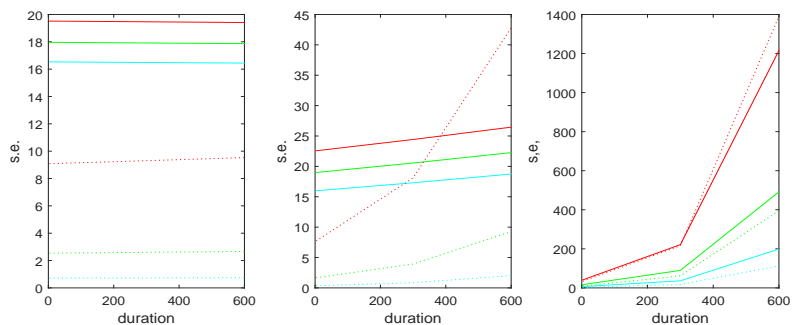}
\caption{\label{Fig2} Plots of $\hat\sigma_{ke}$ (solid lines) and $\hat\tau_{ke}$ (dotted lines) versus duration, separately for each temperature; ordinary pixels are on the left panel, moderately hot pixels on the middle panel and very hot piels in the right panel. }
\end{figure}

Estimates of $\sigma_{ke}$ and $\tau_{ke}$ are displayed together in Figure \ref{Fig2} as function of duration, again separately for each temperature. At -10\degree C and short duration, the standard error specific of individual images is slightly above 16 which is the read-out noise of this camera. Among ordinary pixels the lack of uniformity in the sensor is negligible. It is also small among moderately hot pixels, but is very sensitive to duration and temperature and in bad conditions becomes the main source of variability. The situation among very hot pixels is even more extreme, though these estimates are not sufficiently reliable because the sample size here is just about 19.
Among ordinary pixels the effect of duration seems  negligible on both components of variance while that of temperature seems a little more substantial.  Among moderately hot pixels the effect of duration on image specific errors becomes non negligible but the stronger effect appears to be on the lack of uniformity of the sensor.

\begin{figure}[ht]
\centering
\includegraphics[width=\textwidth,height=4.5cm]{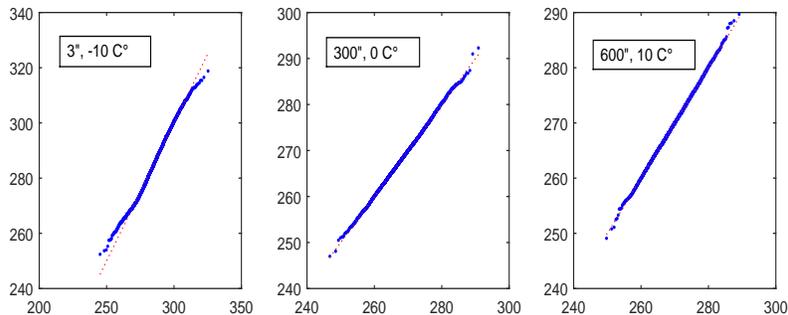}
\caption{\label{Fig3} Quantile/quantile plot of sample averages for ordinary pixels under three different  experimental conditions.}
\end{figure}

A set of quantile/quantile plots of averages among ordinary pixels conditionally on experimental conditions are displayed in Figure \ref{Fig3} to provide an informal critical assessment of the assumption of normality, at least for the sample averages of each set of 10 observations. Discrepancies can be detected mainly at 10\degree C, that is when dark current becomes substantial.
\section{Discussion}
The quality of the data produced by imaging sensors are affected by dark current and hot pixels which introduce bias and additional noise. The data analysed in this paper were acquired with a CCD device for astronomical imaging according to an experimental design aimed at studying the effect of temperature and duration of exposure. A finite mixture model fitted to the data led to detect, besides ordinary and really hot pixels, accounting for about 0.19\% of the total, an intermediate category of moderately hot pixels (about 1.3\% of the total) whose behaviour is deviant but not so extreme like the very hot pixels.

In the analysis it was assumed that, within each latent type, the random effect associated with each pixel is amplified by experimental condition which, however, do not affect the sign and the relative size of these random effects. This conjecture is supported by the results of the experiments in \cite{Burger2011}: they take 200 images under different experimental conditions and show that the ordering of pixels based on the average is not affected by experimental conditions. To check these findings with the Atik camera, 400 images of 3" at 8\degree C and of 2" at 12\degree C were taken and a random sample of 100,000 pixels was selected. For the subset of pixels which at 8\degree C have an average below 360 (99.982\% of the total), a linear regression line was fitted to the averages at 12\degree C with respect to the averages at 8\degree C. The regression coefficient is equal to 1.83 and the standard deviation of the residuals is equal to 2.3; these results seem to indicate that the lack of uniformity is indeed preserved and amplified by changing experimental conditions.

On the whole, our results indicate that both temperature and duration of exposure have a substantial effect on the mean behaviour and noise of hot pixels. However, recent developments in the acquisition and processing of astronomical images, based on applying small random shifts to the camera between images, so that a given hot pixels does not appear in the same position across images; this, combined with a suitable procedure for outlier rejection, can make hot pixels almost irrelevant.

The range of temperatures used in the experiment were limited by the fact that the Atik camera can, at most, achieve -27\degree C below the ambient temperature. An additional limitation in the range of experimental conditions is that, even with only 10 different experimental conditions, the whole dataset was difficult to handle; because of this, analysis was restricted to a random sample of pixels and replications.

Some of the diagnostic plots in Figure \ref{Fig3} indicate that the normal distribution may not provide an adequate approximation of the distribution of the response variable under certain experimental conditions. Though this happens mainly among hot pixels which account for a small minority of the observations, it may have affected certain results. A feasible alternative might be fitting mixtures of skew normal distribution \citep[see]{lee2013}, but the implementation of these methods will require a considerable amount of additional work.

In recent years, CMOS technology is emerging as a possible improvement relative to CCD sensors. It is well known that also CMOS sensors suffer from dark current and hot pixels, however, to the best of our knowledge, no systematic investigation comparable to the one presented in this paper has been conducted.

\bibliographystyle{apalike}
\bibliography{DaCu}
\end{document}